\documentclass[sigconf,natbib=true,anonymous=false,review=false]{acmart}
\usepackage{multicol}
\usepackage{multirow}
\usepackage{amsfonts}
\usepackage{bm}
\usepackage{amsmath}
\usepackage{amsthm}
\usepackage{booktabs} 
\usepackage{acronym}
\usepackage{tcolorbox}
\usepackage{algorithm}
\usepackage{algorithmic}
\usepackage{fancyvrb}
\usepackage{graphicx}
\usepackage{makecell}
\usepackage{hhline}
\usepackage{hyperref}
\usepackage[para]{footmisc}
\usepackage[inline]{enumitem}

\DefineVerbatimEnvironment{MyVerbatim}{Verbatim}
{fontsize=\normalsize, fontfamily=ptm,
 codes={\catcode`\:=12 }, 
 commandchars=\\\{\}} 

\newcommand{\ours}{\text{APCIR}}

\AtBeginDocument{%
  \providecommand\BibTeX{{%
    \normalfont B\kern-0.5em{\scshape i\kern-0.25em b}\kern-0.8em\TeX}}}

\copyrightyear{2025}
\acmYear{2025}
\setcopyright{rightsretained}
\acmConference[CIKM '25]{Proceedings of the 34th ACM International Conference on Management}{November 10--14, 2025}{Seoul, Korea}
\acmBooktitle{Proceedings of the 34th ACM International Conference on Management (CIKM '25), November 10--14, 2025, Seoul, Korea}
\acmDOI{}
\acmISBN{}

\begin{document}

\title{Adaptive Personalized Conversational Information Retrieval}

\author{Fengran Mo}
\orcid{0000-0002-0838-6994}
\authornote{Equal contribution}
\affiliation{%
  \institution{Université de Montréal}
  \city{Montréal}
  \state{Québec}
  \country{Canada}
}
\email{fengran.mo@umontreal.ca}

\author{Yuchen Hui}
\orcid{0000-0002-9659-3714}
\authornotemark[1]
\affiliation{%
  \institution{Université de Montréal}
  \city{Montréal}
  \state{Québec}
  \country{Canada}
}
\email{yuchen.hui@umontreal.ca}

\author{Yuxing Tian}
\orcid{}
\affiliation{%
  \institution{Université de Montréal}
  \city{Montréal}
  \state{Québec}
  \country{Canada}
}
\email{yuxing.tian@umontreal.ca}

\author{Zhaoxuan Tan}
\orcid{}
\affiliation{%
  \institution{University of Notre Dame}
  \city{Notre Dame}
  \state{Indiana}
  \country{USA}
}
\email{ztan3@nd.edu}

\author{Chuan Meng} 
\orcid{0000-0002-1434-7596}
\affiliation{%
  \institution{University of Amsterdam}
  \country{The Netherlands}
}
\email{c.meng@uva.nl}

\author{Zhan Su}
\orcid{}
\affiliation{%
  \institution{Université de Montréal}
  \city{Montréal}
  \state{Québec}
  \country{Canada}
}
\email{zhan.su@umontreal}

\author{Kaiyu Huang}
\orcid{0000-0001-6779-1810}
\authornote{Corresponding author}
\affiliation{%
  \institution{Beijing Jiaotong University}
  \city{Beijing}
  \country{China}}
\email{kyhuang@bjtu.edu.cn}

\author{Jian-Yun Nie}
\orcid{0000-0003-1556-3335}
\authornotemark[2]
\affiliation{%
  \institution{Université de Montréal}
  \city{Montréal}
  \state{Québec}
  \country{Canada}
}
\email{nie@iro.umontreal.ca}

\renewcommand{\shortauthors}{Fengran Mo et al.}

\begin{abstract}
Personalized conversational information retrieval (CIR) systems aim to satisfy users' complex information needs through multi-turn interactions by considering user profiles. However, not all search queries require personalization. 
The challenge lies in appropriately incorporating personalization elements into search when needed.
Most existing studies implicitly incorporate users' personal information and conversational context using large language models without distinguishing the specific requirements for each query turn.
Such a ``one-size-fits-all'' personalization strategy might lead to sub-optimal results.
In this paper, we propose an adaptive personalization method, in which we first identify the required personalization level for a query and integrate personalized queries with other query reformulations to produce various enhanced queries.
Then, we design a personalization-aware ranking fusion approach to assign fusion weights dynamically to different reformulated queries, depending on the required personalization level. 
The proposed \underline{\textbf{a}}daptive \underline{\textbf{p}}ersonalized \underline{\textbf{c}}onversational \underline{\textbf{i}}nformation \underline{\textbf{r}}etrieval framework \textbf{APCIR} is evaluated on two TREC iKAT datasets.
The results confirm the effectiveness of adaptive personalization of APCIR by outperforming state-of-the-art methods.
\end{abstract}





\maketitle


\section{Introduction}
Personalized conversational information retrieval (CIR) systems aim to produce search results to satisfy users' complex information needs based on users' profiles through multi-turn interactions. 
Personalized CIR assumes that the same query from different users may correspond to different search intents, thus yielding different results
~\citep{Zamani2022ConversationalIS,gao2022neural,aliannejadi2024trec,mo2024leverage,abbasiantaeb2024generate,lupart2024irlab,mo2025conversational}.

\begin{figure}[t]
\centering
\includegraphics[width=1\linewidth]{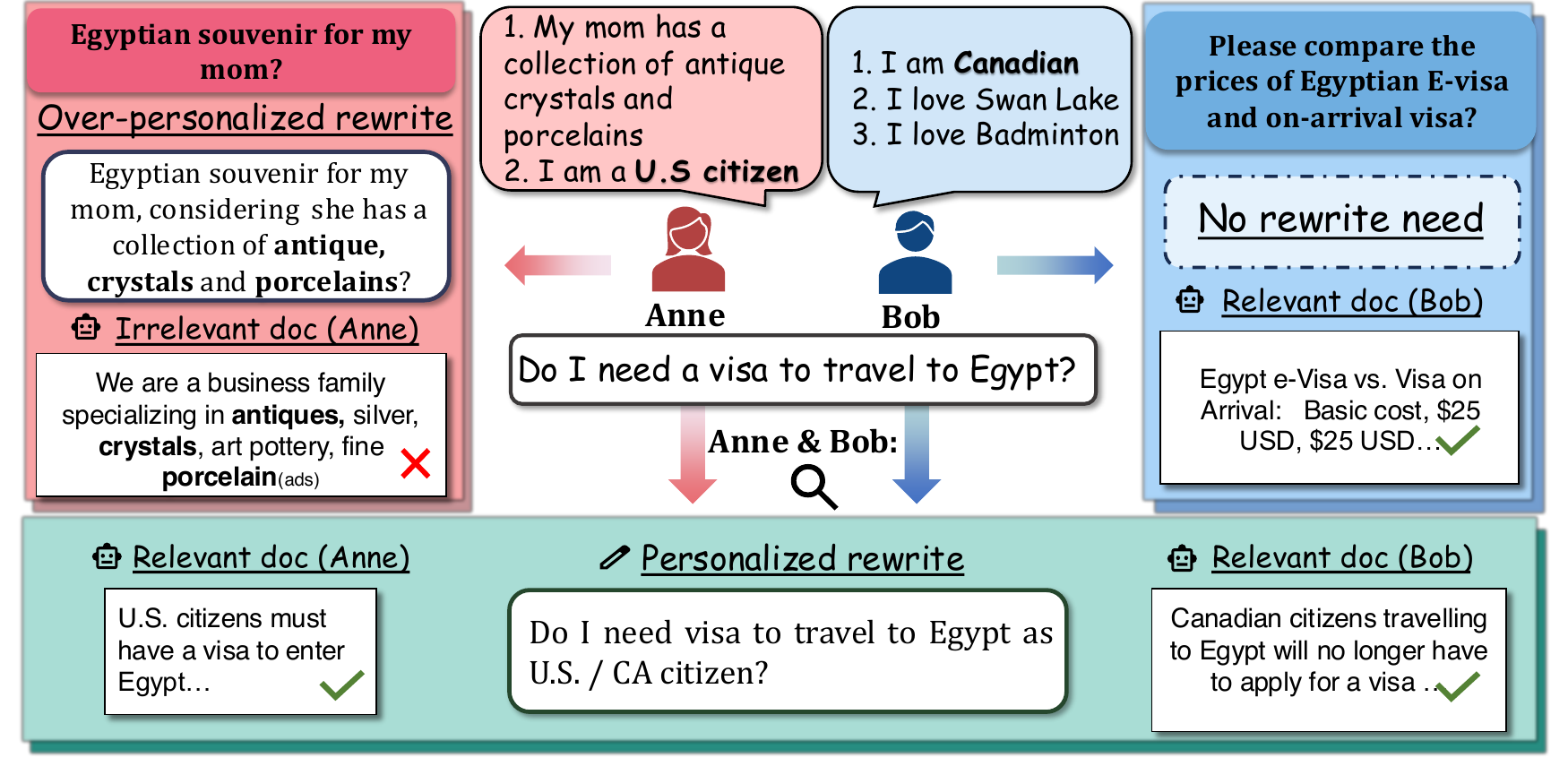}
\vspace{-4ex}
\caption{Example of different personalized information needs for users' queries by considering their profiles.}
\label{fig: example}
\vspace{-4ex}
\end{figure}

The challenge of building a personalized CIR system lies in the appropriate incorporation of useful personalization elements to retrieve relevant information in each query turn~\cite{aliannejadi2024trec,mo2024leverage,abbasiantaeb2024generate,lupart2024irlab}. 
In the example illustrated in Figure~\ref{fig: example}, 
different users' queries require different levels of personalization. 
On the one hand, for a specific query related to personalization, e.g., ``Do I need a visa to travel to Egypt?'', the system needs to identify the relevant information in their personas, such as their citizenship, and build appropriate 
personalized reformulated queries accordingly. 
On the other hand, for a general query such as ``Please compare the prices of Egyptian E-visa and on-arrival visa?'', the system is expected to detect that personalization is unnecessary, and return similar search results for all users.
In many existing studies on personalized search~\cite{tan2006mining,dou2007large,teevan2008personalize,mei2008entropy,wang2013personalized,liu2024personalize}, it is shown that there are risks to performing excessive personalization, i.e., adding unnecessary or irrelevant personal information, which would hurt the overall search performance. For example, when Anne queries Egyptian souvenirs for her mother, who collects antique crystals and porcelains, adding these terms into the query, as a means of personalization, may lead to irrelevant results (e.g., finding antique porcelains from Egypt, which are not representative souvenir items from Egypt).
However, identifying the required personalization level for a given query is non-trivial. 
Unlike the traditional personalized search that can extract user preferences from a large amount of query logs in search engines and use them to train a query personalization classifier, personalized CIR does not have such large training data available. 
This makes it even more challenging to design a method for personalized CIR.

As part of the goal in this paper, we target the problem of detecting the appropriate level of personalization for CIR. We define three levels: personalization, partial personalization, and non-personalization. 
We propose to use large language models (LLMs) to determine the personalization level. 
According to the predicted personalization level, an adaptive ranking fusion method is used to fuse the search results of different reformulated queries, resulting in more appropriate personalized results. 

This approach differs from the common approaches in the literature on personalized CIR, which leverage conversational query reformulation techniques to transform context-dependent queries into stand-alone queries and incorporate personal information from user profiles. 
Among them, \citet{mo2024leverage} conduct relevant user information selection and personalized query reformulation simultaneously by in-context learning based on LLMs. 
Subsequently, an LLM-based generate-then-retrieve framework~\cite{abbasiantaeb2024generate} is proposed to first generate a potential answer with the given user profile and conversational context and then produce a personalized reformulated query grounded on it. 
Recently, an MQ4CS framework designed by \citet{lupart2024irlab} performs multiple query reformulations conditioned on the user profile, and then aggregates the resulting ranking lists using a re-ranker.
However, these approaches treat all query turns with the same personalization strategy by leveraging LLMs to generate a query rewrite that could automatically incorporate some personal information. 
Such an ``one-size-fits-all'' personalization strategy might lead to sub-optimal results, e.g., over-personalization.

The explicit case-by-case detection of the appropriate level of personalization can help solve the above problem. 
Depending on the identified personalization level, the results of the personalized query reformulation will be fused with higher or lower weights, thus helping us to avoid over-personalization when personalization is not required.
We will show in Sec.~\ref{sec: exploration} that such an approach is advantageous compared to non-personalized and uniform personalization approaches. 

More specifically, we develop an \underline{\textbf{a}}daptive \underline{\textbf{p}}ersonalized \underline{\textbf{c}}onversati\-onal \underline{\textbf{i}}nformation \underline{\textbf{r}}etrieval framework \textbf{\ours}, which can identify the required level for personalization. 
\ours~consists of three main components: explicit personalization level identification, different types of query reformulation, and a personalization-aware ranking fusion mechanism. 
The first component aims to determine the required level of personalization using LLMs. The Chain-of-Thought guidance and in-context learning examples are provided to enhance LLMs' predictions. 
The resulting level of personalization is then employed to guide the subsequent conversational query reformulation to generate various types of queries.
The explicit guidance makes the personalization procedure more controllable compared with blindly relying on LLMs to handle all information.
Besides, the LLMs could extract more or less personal information from the user profile according to this explicit signal, resulting in more appropriate personalized queries.
With the various reformulated queries, a personalization-aware ranking fusion mechanism that considers the impact of different personalization levels 
is designed to fuse multiple queries to cover various aspects (both personalized and non-personalized) with adapted weights.
The weights are dynamically assigned to each query turn, making the process different from the traditional fusion with fixed, empirically defined weights for sources.
The personalization-aware ranking fusion can also alleviate the possible inaccurate personalization level identification and potential query drift caused by personalization. 

Experiments are conducted on the datasets of Interactive Knowledge Assistance Track (iKAT) of Text Retrieval Conference (TREC)~\cite{aliannejadi2024trec,abbasiantaeb2025conversational}, which offers the first test collections for personalized CIR. An explicit user profile with a set of natural language sentences is available to describe various characteristics or preferences of a user (e.g., ``I am Canadian'' in Figure~\ref{fig: example}).
A user's profile may contain different aspects, which might or might not be related to a specific query at a turn. 
Thus, a desirable personalized CIR system should be able to perform context-dependent personalization 
conditional on the historical context and user profiles.
We demonstrate that our \textbf{\ours}~ achieves higher performance on these datasets than the existing state-of-the-art baseline methods, and our devised dynamic personalized weight identification mechanism for retrieval fusion is more effective than the other personalization approaches.
Detailed ablation analyses show the contributions of different components in our adaptive personalized CIR approach.

Our contributions are summarized as follows:
(1) We propose an LLM-based framework to explicitly determine the appropriate level of personalization and integrate it with the user profile for query reformulation in CIR. 
(2) We design a personalization-aware ranking fusion mechanism to achieve adaptive personalized conversational information retrieval based on dynamic weight assignment. Personalized search results are incorporated into the final results depending on the required personalization level, thus helping to avoid problems with a uniform fusion method that may lead to over-personalization. 
(3) Experiments on two TREC iKAT datasets demonstrate the superior performance of our framework by outperforming all existing baselines. 
We provide our code and data at our released~\href{https://github.com/YuchenHui22314/TREC_iKAT_2024}{Github Repository}. 




\section{Related Work}
\label{sec: Related Work}
\noindent \textbf{Conversational Information Retrieval} (CIR)~\cite{qu2020open,kim2022saving,meng2023query,mo2024survey}
enables users to interact with a search system via multi-turn conversations to address their complex information needs. 
The challenge of CIR mainly lies in understanding users' real search intent.
Two main research lines in the literature are conducted for CIR: (i) building a conversational dense retriever (CDR) and (ii) conversational query rewriting (CQR).
The CDR methods~\cite{yu2021few,lin2021contextualized,mao2022curriculum,jin2023instructor,cheng2024interpreting,mao2024chatretriever,lupart2024disco,mo2025uniconv} directly encode the whole conversational search session into the model to perform end-to-end dense retrieval
~\cite{kim2022saving,mo2024history}. 
This paradigm highly relies on the quantity and quality of the training samples with relevance judgment~\cite{dai2022painting,mao2022convtrans,mo2024convsdg,chen2024generalizing,meng2025query,mo2025convmix}. Since conversational search engines are not widely deployed, the scarcity of available data for model development is challenging.
On the other hand, the CQR methods aim to convert the context-dependent queries into stand-alone ones. Then, any existing ad-hoc search retrievers can be applied to the reformulated query following the rewrite-then-retrieval pipeline~\cite{elgohary2019can,meng2025bridging}. This approach demonstrated substantial practical value for conversational search in low-resource scenarios.  
Earlier CQR studies attempted to select useful tokens from the conversation context~\cite{2020Making,voskarides2020query,fang2022open} or train a query rewriter with conversational sessions to mimic the human rewrites~\cite{yu2020few,lin2020conversational,vakulenko2021question}.
To mitigate the discrepancy between rewriter training and the downstream search task, some studies adopt reinforcement learning~\cite{wu2022conqrr,chen2022RLCQR} or exploit the ranking signals with the rewriting model training~\cite{mo2023convgqr,mao2023search}, while others jointly learn query rewriting and context modeling~\cite{qian2022explicit,mo2023learning}.
With the emergence of LLMs, recent methods adapt LLMs to directly generate the query rewrites for retrieval~\cite{ye2023enhancing,mao2023large} or to generate high-quality pseudo queries as supervision signals for rewrite model training~\cite{jang2023itercqr,yoon2024ask,lai2025adacqr,zhang2024adaptive,mo2024chiq}. 
Our approaches follow the CQR paradigm for personalization in CIR.

\noindent \textbf{Personalized Conversational Information Retrieval.}
Personalized search~\cite{speretta2005personalized,micarelli2007personalized,zhang2024personalization,xu2025personalized} adapts the models with user-specific contexts to cater to the unique information needs and helps the system avoid providing identical search results for all users without considering individual preferences. User preferences can be collected from query logs in traditional ad-hoc search scenarios~\citep{mei2008entropy,wang2013personalized,zhou2020encoding,liu2024personalize}. 
However, the personalized search models based on deep neural networks~\cite{zhou2024cognitive} require abundant training data, which is unavailable in practical CIR scenarios. 
In addition, incorporating user profiles for personalized CIR~\cite{aliannejadi2024trec,abbasiantaeb2025conversational} becomes more complex because each user's background may contain different aspects, which might not all be related to the query in the current turn. 
Either excessive personalization or non-personalization would hurt the overall search performance if the model cannot adaptively identify the appropriate level of personalization~\cite{dou2007large,teevan2008personalize,mo2024leverage}.
Existing studies investigated how to leverage LLMs to select the relevant information from a user profile for each query turn~\cite{patwardhan2023sequencing} or incorporate it into query reformulation~\cite{mo2024leverage}. 
Different from them, we perform adaptive personalized CIR by explicitly identifying the relevant personal information for each query turn and using it to generate various reformulated queries for retrieval fusion with dynamic personalized weights.


\section{Personalization in Conversational IR}
\subsection{Task Definition}
Personalized conversational IR aims to retrieve relevant passages $p^+$ from a collection $\mathcal{C}$ to satisfy the information need of the current query turn $q_n$ of a specific user, given the historical conversational context $\mathcal{H}_n =\{q_i,r_i\}_{i=1}^{n-1}$ and the user profile $\mathcal{U}$~\cite{aliannejadi2024trec,abbasiantaeb2025conversational}, where $q_i$ and $r_i$ denote the query and response of $i$-th turn, and $\mathcal{U}$ consists of a set of natural language sentences that describe the user background or preferences. 
The current query turn $q_n$ is context-dependent and might conceal specific personalized information needs. 
Our goal is to make adaptive personalization in search by leveraging relevant personalization elements and conversational context. 

\subsection{Exploration of Fusion for Personalization}
\label{sec: exploration}
The general assumption behind the existing query reformulation methods for personalized CIR~\cite{mo2024leverage,abbasiantaeb2024generate,patwardhan2023sequencing,lupart2024irlab} is that \textit{all query turns require personalization, and that deploying LLMs for query reformulation can implicitly identify and incorporate the useful personalization elements.} 
Therefore, in the previous approaches, the whole user profile is injected into the context of the current turn's query reformulation, in a similar way to the conversation context.
Although they improve performance in some query turns compared to reformulating queries without user information, this ``one-size-fits-all'' strategy may fail to identify all the useful personal information or perform personalization when this is unnecessary. 
Indeed, not all query turns in a conversational session need to be personalized. This is also observed in some earlier studies on ad-hoc personalized search~\cite{dou2007large,teevan2008personalize} -- personalizing some queries may lead to decreased effectiveness. 
On the other hand, for those query turns that need to be personalized, the explicit user information selection~\cite{aliannejadi2024trec} might lead to information loss due to the potential error in binary judgment, i.e., determining whether a sentence/aspect in the user profile should be used for query reformulation.

\begin{figure}[t]
\centering
\includegraphics[width=1\linewidth]{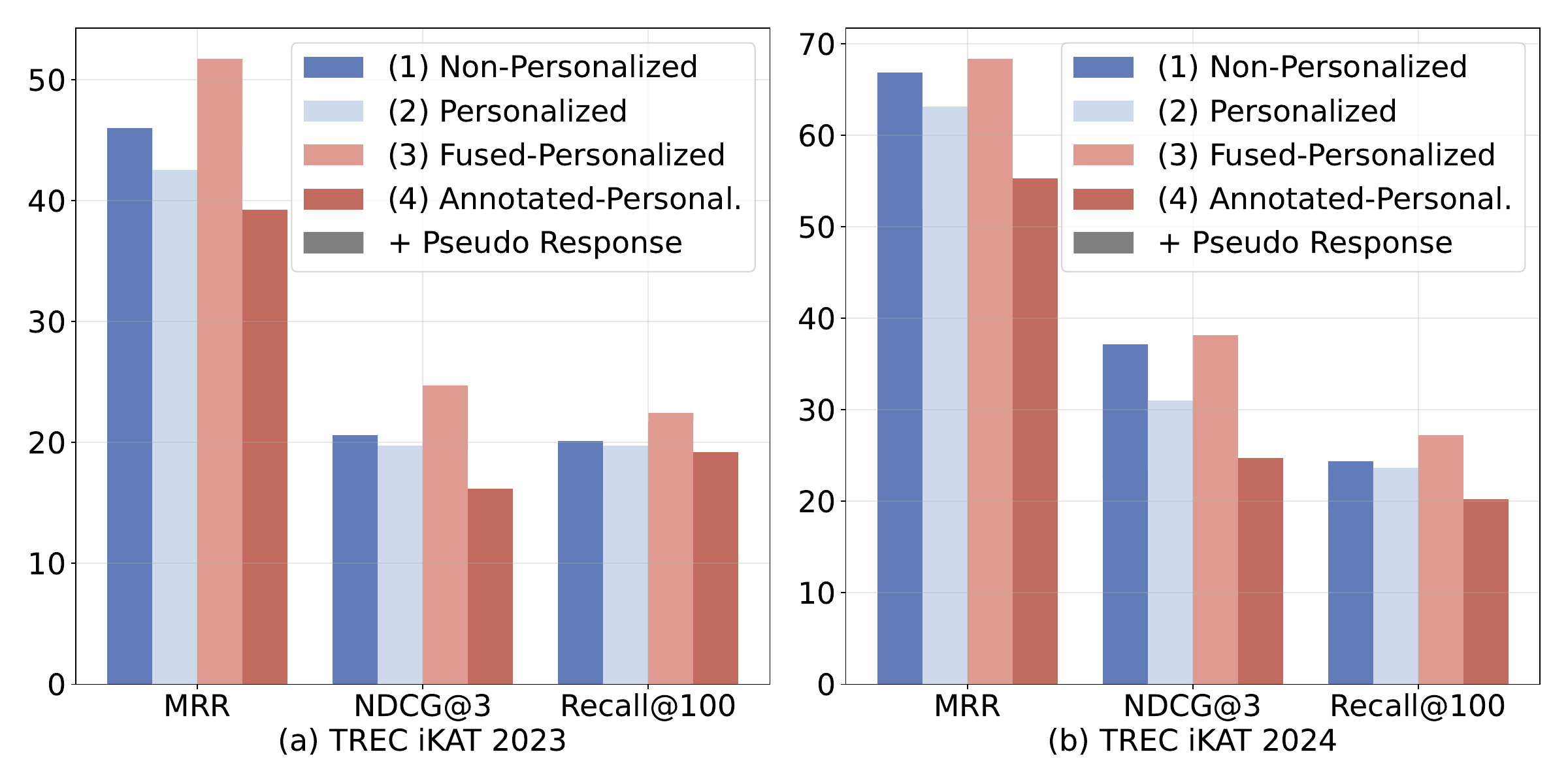}
\vspace{-4ex}
\caption{Preliminary experimental results with four different scenarios on two TREC iKAT datasets.}
\label{fig: preliminary}
\vspace{-4ex}
\end{figure}

To verify how personalization impacts query reformulation in CIR, we run a preliminary experiment to compare the following scenarios: (1) Reformulating queries based on conversational context without user profiles (Non-Personalized), (2) Reformulating queries based on user profile and conversational context for personalization (Personalized), (3) Combining the search results of the previous two via ranking fusion by linearly aggregating different ranking lists based on their relevance scores with a default weight of $1$  following~\cite{lin2021contextualized} (Fused-Personalized), and (4) Conducting personalized and non-personalized query reformulation according to binary human annotations on the relevance of each sentence/aspect in the user profile w.r.t the current turn query (Annotated-Personalized). 
In addition, we also validate the effectiveness of leveraging pseudo responses generated by LLMs for query expansion~\cite{wang2023query2doc}.
Notice that method (2) is widely used in existing work, and method (4) should be based on ground-truth annotations, which is available in the test collection but unrealistic in real situations.
We want to re-examine the ability of LLMs to accurately leverage the user information for personalization and show that retrieval result fusion is a potential 
strategy to achieve more adaptive personalization, 
compared to relying on LLMs solely or even with identifiable annotations.

\begin{figure*}[t]
\centering
\includegraphics[width=1\linewidth]{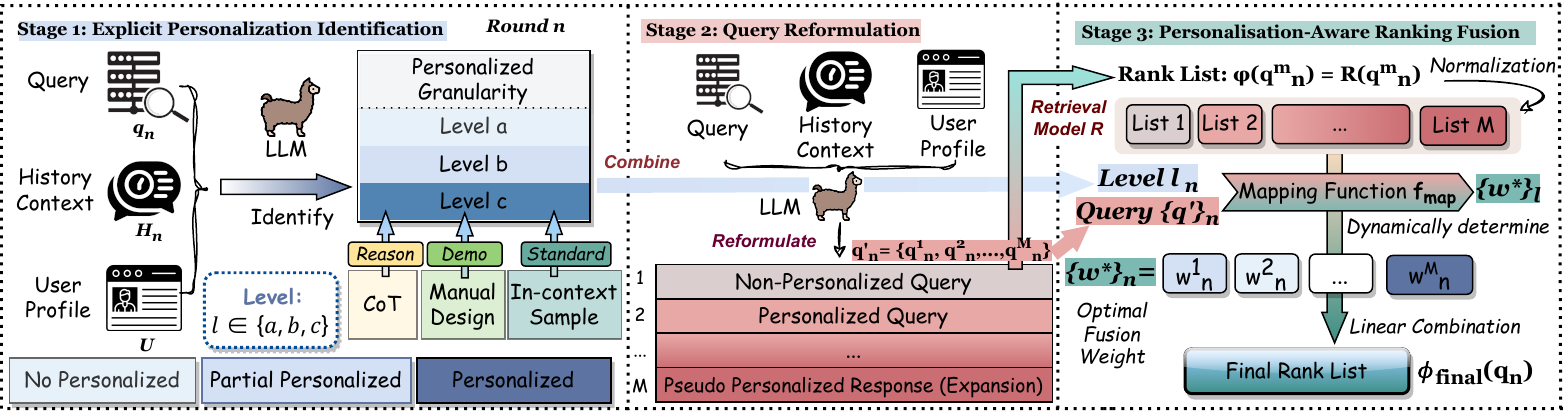}
\vspace{-4ex}
\caption{Overview of our \ours~framework and workflow. The personalization granularity is first identified for each given query (Stage 1) and then used to reformulate multiple queries (Stage 2). The personalization level and reformulated queries are provided for personalization-aware ranking fusion (Stage 3) to obtain optimal fusion weights and produce a final ranking list.}
\label{fig: overview}
\vspace{-3ex}
\end{figure*}

We employ GPT-4o~\cite{gpt4o} and BM25~\cite{robertson1995okapi} models for query reformulation and retrieval on the two TREC iKAT datasets, respectively. 
The personalized reformulated queries $q_n^{u}$, the non-personalized one $q_n^{\prime}$, as well as their respective pseudo response $r_n^{u}$ and $r_n^{\prime}$ are generated with or without user profile $\mathcal{U}$ with instruction $\mathcal{I}$ as 
\raggedbottom
\begin{equation}
\label{eq: pre}
    q_n^{u}, r_n^u = \mathcal{LLM}(\mathcal{U}, \mathcal{H}_n \circ q_n, \mathcal{I}), \quad q_n^{\prime}, r_n^{\prime} = \mathcal{LLM}(\mathcal{H}_n \circ q_n, \mathcal{I})
\end{equation}

The experimental results are presented in Figure~\ref{fig: preliminary}. We observe that personalized reformulated queries do not improve the overall retrieval performance compared to the non-personalized ones, while fusing these two types (Fused-Personalized) obtains better results by aggregating their relevance scores in different ranking lists.
This suggests that simply including or discarding user profiles for query reformulation cannot adequately enhance the query with the relevant personal information, while ranking fusion has the potential to achieve it. 
We also notice that relying on human identification of profile relevance (Annotated-Personalized) does not improve the results, while using the generated pseudo responses is generally beneficial for retrieval. 
The reason might be that a binary relevance annotation is not precise enough, while the parametric knowledge in LLMs is helpful.
In summary, the above preliminary tests indicate that an adaptive fusion method using both personalized and non-personalized search results according to the need could be a better approach than a uniform personalization method.

\section{Methodology}
As shown in the previous section, using personalized queries formulated by an LLM with full access to all user information might result in much worse performance than the non-personalized approach due to possible over-personalization. 
In addition, even with human-selected personal information, the LLM-reformulated query still cannot perform better than the ranking fusion approach.

To achieve adaptive personalization for CIR, our method performs an explicit personalization level identification, then leverages the identified level to reformulate context-dependent queries into different types, and finally fuses them.
However, we also notice that the personalization level identified by LLMs is difficult to transform into concrete weights for specific fusion mechanisms without corresponding training. 
Due to the scarcity of training data, we design a training-free personalization-aware ranking list fusion mechanism to determine the fusion weights for personalized and non-personalized search results. 
The overview of our method is depicted in Figure~\ref{fig: overview}, which consists of three main components: explicit personalization level identification, various query reformulations for personalized CIR, and personalization-aware ranking fusion. 
Our framework can be integrated with any query reformulation models, retrievers, and retrieval results fusion strategies.

\subsection{Explicit Personalization Level Identification}
\label{sec: Explicit Personalization Identification}
Explicit personalization level identification aims to obtain the required personalization level for the current query. 
This identified level can help reduce the difficulty for LLMs in de-noising irrelevant information in user profiles for the current query and make personalization more controllable and explainable, i.e., based on the judgments from LLMs, the system can determine to what extent personalization should be applied to the query. 
Since no data is available to support training a model for personalization level identification, we achieve it by instructing LLM with in-context learning and chain-of-thought (CoT) reasoning.
Formally, we ask the LLMs to identify the appropriate personalization level for the current turn by providing the historical context $\mathcal{H}_n$, the current query $q_n$, and the corresponding user information $\mathcal{U}$ with crafted instructions $\mathcal{I}$.

We first define three levels of personalization, incorporated as part of the instruction to LLMs: (a) non-personalization -- the self-contained query that does not need personalization; (b) partial personalization -- the user profile serves as an extra perk, while the query itself can retrieve some general topics; and (c) personalization -- the user profile presents important and indispensable information or constraints for an accurate answer. 
LLMs are instructed to provide one of these levels as output for a turn. 
Since LLMs are not optimized for personalization~\cite{zhang2024personalization}, we also design examples $L_\text{Example}$ to illustrate the levels concretely, together with hand-crafted demonstrations for in-context learning to help the LLMs understand the required task for a specific query.
In addition, to further arouse the reasoning ability of LLMs, we implement a CoT mechanism to elicit reasoning from LLMs before identifying level and rewriting query. 
This procedure is formulated as
\raggedbottom
\begin{equation}
    q_n^{\text{level}} = \mathcal{LLM}(\mathcal{U}, \mathcal{H}_n \circ q_n, \mathcal{I} \ | \ L_\text{Example}, \text{CoT})
\end{equation}
\raggedbottom
\subsection{Query Reformulation for Personalized CIR}
\label{sec: Personalized Query Reformulation}
We adopt query reformulation approaches to integrate personalization elements with conversational context due to the data scarcity issue~\cite{mao2022convtrans,chen2024generalizing}. 
The principle is that when an LLM is informed of the personalization level, it can try to extract more or less personal information from the user profile according to it, resulting in a more appropriate personalized query.
With a CoT mechanism, we could further ask LLM to reason in two steps: first, identify the level, then generate a query.

Specifically, given the identified level from the previous section, if it requires personalization, we leverage LLMs to reformulate the context-dependent query $q_n$ into a stand-alone one $q_n^u$ and generate the corresponding pseudo response $r_n^u$ for personalized CIR together with the user profile $\mathcal{U}$ as Eq.~\ref{Eq: per_QR}. Otherwise, if personalization is unnecessary, a general reformulation to generate any number of rewrites without user profile $\mathcal{U}$ is applied as Eq.~\ref{Eq: general_QR}.
\raggedbottom
\begin{equation}
\label{Eq: per_QR}
    q_n^{u}, r_n^{u} = \mathcal{LLM}(\mathcal{U}, \mathcal{H}_n \circ q_n, \mathcal{I} \ | \ q_n^{\text{level}}\in \text{Per.}, QR_\text{Exam}, \text{CoT})
\end{equation}
\begin{equation}
\label{Eq: general_QR}
    \{q_n^{\prime}, r_n^{\prime}\} = \mathcal{LLM}(\mathcal{H}_n \circ q_n, \mathcal{I} \ | \ q_n^{\text{level}}, QR_\text{Example}, \text{CoT})
\end{equation}
where the CoT mechanism and the in-context personalized query reformulation example $QR_\text{Example}$ are also provided to illustrate how the de-contextualization is adapted for personalization.

The non-personalized reformulated query $q_n^{\prime}$ and its pseudo response $r_n^{\prime}$ will later be employed in ranking fusion to prevent query drift and over-personalization. 
When no personalization is needed ($q_n^{\text{level}}\notin \text{Personalized}$), another type of reformulated query $q_n^{\prime\prime} \circ r_n^{\prime\prime}$ is generated as Eq.~\ref{Eq: general_QR} to ensure the number of reformulated query variants is equal to the case that the requiring personalization ($q_n^{\text{level}} \in \text{Personalized}$), which can be flexibly controlled by adjusting the instruction $\mathcal{I}$.
In our implementation, this query reformulation stage is integrated with that of personalization level identification of the previous section and produces reformulated queries simultaneously to reduce the inference latency by calling the LLMs only once.




\subsection{Personalization-Aware Ranking Fusion}
\label{sec: Fine-Grained Retrieval Fusion}
In Sec.~\ref{sec: exploration}, we observed that ranking fusion helps to obtain better personalized CIR results. 
The common practice for fusing different queries 
is to combine their associated ranking lists (e.g., round-robin fusion) or representations (e.g., summing up the representations).
However, one usually assigns fixed fusion weights to each source~\cite{kurland2018fusion,lin2021contextualized,huang2024unleashing}. This strategy is not suitable for adaptive personalization that requires taking into account the specific personalization needs dynamically. 
Thus, we design a personalization-aware ranking fusion mechanism to leverage the various types of reformulated queries 
for adaptive personalized CIR.

\begin{algorithm} 
    \caption{Adaptive Personalized Conversational Information Retrieval with Personalization-Aware Ranking Fusion}
    \label{alo: fine-grained-fusion} 
    \renewcommand{\algorithmicrequire}{\textbf{Input:}}
    \renewcommand{\algorithmicensure}{\textbf{Output:}}
    \begin{algorithmic}[1]
    \REQUIRE  A conversation session with $N$ query turns $\{\mathcal{Q}\}_N$ and associated user profile $\mathcal{U}$, current query turn $q_n$ and its corresponding history context $\mathcal{H}_n$. A retriever model $\mathcal{R}$ and a evaluation metric $\mathcal{M}$, and a linear combined fusion function $\mathcal{F}_{\text{lc}}(\cdot)$.
    \ENSURE The final personalized ranking list $\phi_{\text{final}}(\{\mathcal{Q}\}_N)$.\\ 
    \STATE \textit{Perform explicit personalization level identification (Sec.~\ref{sec: Explicit Personalization Identification}) and personalized conversational query reformulation (Sec.~\ref{sec: Personalized Query Reformulation})}
    \STATE Initialize lists for each personalization level: $L_a, L_b, L_c \gets []$  
    \FOR {\textbf{each} conversational query turn $q_n \in \{\mathcal{Q}\}_N$}
        \STATE Reformulate query and identify the personalization level $l_n$
        \STATE $l_n, q_n^{u}, r_n^{u} \gets \mathcal{LLM}(\mathcal{U}, \mathcal{H}_n \circ q_n, \mathcal{I}), \text{ if } l_n \neq a $
        \STATE $q_n^{\prime}, r_n^{\prime} (\{\text{or }q_n^{\prime}, r_n^{\prime}, q_n^{\prime\prime}, r_n^{\prime\prime}\} \text{ if } l_n == a) \gets \mathcal{LLM}
        (\mathcal{H}_n \circ q_n, \mathcal{I})$
        \FOR {$QR$ \textbf{in} [$q_n^{\prime}, q_n^{\prime}+r_n^{\prime},  q_n^{u}+r_n^u/q_n^{\prime\prime}+ r_n^{\prime\prime}]$} 
            \STATE Perform retrieval and score normalization as
            \STATE 
            RankingList $\phi(QR) \gets \mathcal{R}(QR), \ \phi(QR) \gets$ $f_{\text{norm}}(\phi(QR))$
        \ENDFOR
        \STATE Add $q_n$ to its identified $l_n$ personalization level list $L_{l}$
    \ENDFOR
    \STATE \textit{Perform personalization-aware ranking fusion via reformulated queries $QR$ and personalized levels $l_n$ (Sec.~\ref{sec: Fine-Grained Retrieval Fusion})}
    \FOR {$l$ \textbf{in} personalized level set [a,b,c]}
        \STATE Initialize best score and fusion weight $\mathcal{S}_l\gets0$, $\{w^*\}_l\gets\{\}$
        \FOR {$w_1,w_2,w_3$ s.t. $w_1+w_2+w_3=1$} 
            \STATE Initialize the overall score of the weight set: $\mathcal{S} \gets 0$
            \FOR {\textbf{each} turn $q_n$ \textbf{in} $L_l$}
            \IF {$L_l$ == a}
                \STATE $\phi_{\text{fused}}(q_n)= w_1\phi(q_n^{\prime})+w_2\phi(q_n^{\prime}+r_n^{\prime})+w_3\phi(q_n^{\prime\prime}+r_n^{\prime\prime})$
            \ELSE
                \STATE $\phi_{\text{fused}}(q_n)= w_1\phi(q_n^{\prime})+w_2\phi(q_n^{\prime}+r_n^{\prime})+w_3\phi(q_n^{u}+r_n^u)$
            \ENDIF
            
            \STATE $\mathcal{S} = \mathcal{S} +\mathcal{M}(\phi_{\text{fused}}(q_n))$ 
            \ENDFOR
            \STATE \textbf{If }$\mathcal{S} > \mathcal{S}_l$ \textbf{then} $\mathcal{S}_l \gets \mathcal{S}, \{w^*\}_l \gets \{w_1,w_2,w_3\}$
        \ENDFOR
    \ENDFOR
    \FOR {\textbf{each} conversational query turn $q_n \in \{\mathcal{Q}\}_N$}
        \STATE $ \phi_{\text{final}}(q_n)=\mathcal{F}_{\text{lc}}(\{ w^* \}_{l_n}, \phi(\{q^{\prime}\}_n))$
        \STATE Add $\phi_{\text{final}}(q_n)$ to the final ranking list $\phi_{\text{final}}(\{\mathcal{Q}\}_N)$
    \ENDFOR
    \end{algorithmic} 
\end{algorithm}

Specifically, given a user-system conversation session with $N$ query turns $\{\mathcal{Q}\}_N$, where each turn $q_n$ generates $M$ reformulated queries (including personalized queries) $\{q'\}_n=\{q_n^1, q_n^2..., q_n^m..., q_n^M\}$, $  m\in[1,M]$, the crucial step is to determine a set of fusion weights $\{w\}_n$ to combine the results of the reformulated queries.
The weights should be conditioned on the personalization level $l_n$, rather than using the same set of static weights across all turns.
Intuitively, the set of weights should vary depending on the personalization level, but we assume that they remain consistent across all turns that share the same level of personalization. 
Thus, we expect a mapping function $f_{\text{map}}$ to dynamically determine the optimal fusion weight $\{w^*\}_n = f_{\text{map}}(l_{n}, \{q'\}_n) = \{w_n^1,w_n^2,...,w_n^M\}$ for the reformulated queries at the current turn $\{q'\}_n = \{q_n^1, q_n^2,..., q_n^M\}$. 
Concretely, we first define a set of possible weight values ranging from $0$ to $1$ with the interval of $0.01$ as candidates for each reformulated query in the same turn, and expect the best combination could be selected.
Then, we separate three distinct groups of fusion weights: $\{w\}_a,\{w\}_b$, and $\{w\}_c$ corresponding to the multiple levels of personalization described in Sec.~\ref{sec: Explicit Personalization Identification}. 
Given the ranking list $\phi(q_n^m)=\mathcal{R}(q_n^m)$  obtained by a retrieval model $\mathcal{R}$ for each reformulated query $q_n^m  \in \{q^{\prime}\}_n$ before the weight optimization process,  a min-max normalization is applied to each ranking list to ensure uniform scoring for all retrieved passages on the scale of $[0,1]$ as follows:
\vspace{-1ex}
$$
f_{\text{norm}}(s(q_n^m,p)) = \frac{s(q_n^m, p)- S_{min} }{S_{max}-S_{min}}
\vspace{-1ex}
$$ 
where $s(q_{n}^{m},p)$ is the ranking score between the reformulated query rewrite and a candidate passage, and $S_{max}$ and $S_{min}$ denotes the maximum and minimum passage scores in the ranking list $\phi(q_n^m)$. 

Then, for each level $l \in \{a, b, c\}$, the optimal weights $\{ w^{*} \}_{l}$ are defined as the ones that result in the best search performance for all the query turns in the group as Eq.~\ref{eq: maximum_score}, and  
satisfying 
$\sum_{m =1}^M w_{l}^{m} =1$.
\vspace{-0.8em}
\begin{equation}
\label{eq: maximum_score}
 \arg \max_{\{w\}_l} \sum_{ \substack{n=1}}^{N} \mathcal{M}\left (\mathcal{F}_{\text{lc}}(\{ w \}_{l}, \phi(q_{n}^{1}),\phi(q_{n}^{2}),\dots,\phi(q_{n}^{M})\right ), \ l_{n} = l
\end{equation}
\raggedbottom
where $\mathcal{M}$ and $\mathcal{F}_{\text{lc}}(\{w\},\phi(\cdot))$ denote a retrieval evaluation metric and the linear combination fusion function, respectively. 
The operator of argmax is implemented as a grid search. 
We use a set of validation queries for selecting fusion weights and then apply them to the test queries.
Finally, the final ranking list is produced by linearly combining the search results using the obtained best weights $\{ w^{*} \}_{l}$ for the identified level. 
The whole procedure for achieving adaptive personalized CIR with personalization-aware ranking fusion is described in Algorithm~\ref{alo: fine-grained-fusion}.
\section{Experimental Setup}
\subsection{Dataset and Evaluation Metrics}

\noindent \textbf{\textit{Datasets}.} We conduct experiments on two available personalized CIR datasets, TREC iKAT 2023~\cite{aliannejadi2024trec} and TREC iKAT 2024~\cite{abbasiantaeb2025conversational}. 
Different from other available CIR datasets~\cite{dalton2020trec,dalton2021cast,dalton2022cast,owoicho2022trec,adlakha2022topiocqa,anantha2021open}, the two TREC iKAT datasets introduce a new feature by enhancing the conversations with the background and preferences of the users and adding more complex information needs. 
The user profile is provided as a set of natural language sentences and is static during the conversation. An ideal system needs to do reasoning over context, user profile, and different sources of information to respond to the complex information needs. 
Both datasets share the same document collection, which is a subset of ClueWeb 22B Corpus~\cite{overwijk2022clueweb22}. 

\noindent \textbf{\textit{Evaluation}.} Following previous studies~\cite{abbasiantaeb2024generate,mo2024leverage}, we use standard metrics: MRR, NDCG, and Recall to evaluate the retrieval effectiveness. 
The \textit{pytrec\_eval} tool~\cite{sigir18_pytrec_eval} is used for metric computation.

\begin{table*}[t]
    \centering
    \caption{Performance of methods on two personalized CIR datasets. \textbf{Bold} indicate the best results except 
    \text{HumanQR}. $\dagger$ denotes significant improvements with t-test at $p<0.05$ with Bonferroni correction over the main competitor \text{PCIR}, \text{GtR}, and \text{MQ4CS}.} 
    \vspace{-3ex}
    \begin{tabular}{cclcccccccc}
    \toprule
        \multicolumn{2}{c}{\multirow{2}[2]{*}{Setting}} &  \multicolumn{1}{c}{\multirow{2}[2]{*}{Method}}  & \multicolumn{4}{c}{TREC iKAT-23} & \multicolumn{4}{c}{TREC iKAT-24} \\ 
        \cmidrule(lr){4-7} \cmidrule(lr){8-11}
        & & & MRR & N@3 & R@10  & R@100 & MRR & N@3 & R@10 & R@100 \\ 
        \midrule
        \multirow{11}[7]{*}{Open} &\multirow{7}[5]{*}{Retrieval} & T5QR & 30.2 & 14.1 & 4.4 & 13.4 & 37.6 & 15.4 & 2.7 & 11.2 \\
        & &ConvGQR & 33.9 & 14.7 & 4.0 & 13.1 & 36.4 & 16.0 & 2.5 & 10.8\\
        \multirow{10}[8]{*}{Comparison} & & GPT4QR &  32.2 & 14.1 & 4.5 & 16.1 & 43.2 & 18.1 & 4.1 & 15.7\\
        & &PCIR & 45.7 & 22.7 & 7.1 & 19.8 & - & - & - & -   \\
        & &GtR & 38.6  & 23.0 & 6.8 & \textbf{26.7} & - & - & - & - \\
        \cmidrule(lr){3-11}
        && \ours~(Ours) & \textbf{57.3}$^\dagger$ & \textbf{31.1}$^\dagger$ & \textbf{8.2}$^\dagger$ & 26.2 & \textbf{82.7}$^\dagger$ & \textbf{50.8}$^\dagger$ & \textbf{9.1}$^\dagger$ & \textbf{40.2}$^\dagger$ \\
        \cmidrule(lr){3-11}
        & &HumanQR & 59.8 & 29.6 & 9.8 & 30.6 & 76.8 & 41.5 & 7.9 & 34.6 \\  
        \hhline{~==========}
        &\multirow{4}[2]{*}{+Reranking} &GtR & 42.9 & 18.1 & 4.9 & 21.2 & - & - & - & -\\
        & &MQ4CS & - & - & - & - & 84.8 & \textbf{53.2} & - & \textbf{43.7} \\
        \cmidrule(lr){3-11}
        & &\ours~(Ours) &  \textbf{54.4} & \textbf{28.9} & \textbf{8.3} & \textbf{26.2} & \textbf{85.1} & \textbf{53.2} & \textbf{9.7} & 40.2\\
        \cmidrule(lr){3-11}
        & &HumanQR & 67.5 & 39.3 & 11.2 & 30.6 & 81.9 & 47.6 & 9.4 & 34.6\\
        \toprule
        & \multirow{4}[2]{*}{Retrieval}&PCIR & 47.6 & 24.2 & 7.1 & 22.3 & 63.6 & 29.9 & 6.2 & 25.4\\
        & &GtR   & 40.0 & 18.6 & 5.3 & 18.1 & 61.8 & 28.4 & 5.4 & 22.3\\
        \multirow{5}{*}{Aligned}  &   & MQ4CS & 40.4 & 18.0 & 5.1 & 18.3 & 58.1 & 27.5 & 5.7 & 22.8 \\
        \cmidrule(lr){3-11}
        & &\ours~(Ours) & \textbf{53.6}$^\dagger$ & \textbf{26.6}$^\dagger$ & \textbf{7.6}$^\dagger$ & \textbf{26.2}$^\dagger$ & \textbf{79.7}$^\dagger$ & \textbf{42.1}$^\dagger$ & \textbf{7.5}$^\dagger$ & \textbf{29.0}$^\dagger$ \\
        \hhline{~==========}
        &\multirow{4}[2]{*}{+Reranking} &PCIR & 51.8 & 25.5 & 7.5 & 22.3 & 78.2 & 45.5 & 8.7 & 25.4\\
        & &GtR   & 49.1 & 23.4 & 6.8 & 18.1 & 81.6 & 46.1 & 8.9 & 22.3\\
        & &MQ4CS & 48.4 & 24.2 & 6.5 & 25.3 & 83.6 & 47.2 & 9.1 & \textbf{39.9}\\
        \cmidrule(lr){3-11}
        & &\ours~(Ours) & \textbf{54.4}$^\dagger$ & \textbf{28.9}$^\dagger$ & \textbf{8.3}$^\dagger$ & \textbf{26.2}$^\dagger$ & \textbf{84.1}$^\dagger$ & \textbf{49.6}$^\dagger$ & \textbf{9.3} & 29.0 \\
        \bottomrule
     \end{tabular}
     \label{table: main_results}
\vspace{-2ex}
\end{table*}
\subsection{Baseline Methods}


We compare our method \ours~ with two categories of baseline methods.
The first category is conversational query reformulation systems in the personalized CIR task, aiming to evaluate the effectiveness of the overall framework.
The second one is personalized degree identification approaches, focusing on evaluating the personalized weight selection mechanisms in ranking fusion.

For the first category,
we mainly compare with the following CQR systems:
(1) \textbf{T5QR}~\cite{lin2020conversational}: A strong T5-based model for query reformulation. (2) \textbf{ConvGQR}~\cite{mo2023convgqr}: Combining two T5-based models for query rewrite and query expansion in query reformulation. (3) \textbf{GPT4QR}: Prompting GPT4 directly for personalized query reformulation with the given $ \mathcal{U}$ and $ \mathcal{H}_n$.
(4) \textbf{PCIR}~\cite{mo2024leverage}: Leveraging 3-shot in-context learning to instruct ChatGPT to generate reformulated query and pseudo response with the given $ \mathcal{U}$ and $ \mathcal{H}_n$.
(5) \textbf{GtR}~\cite{abbasiantaeb2024generate}: Leveraging GPT-4 to generate potential answers with the given user profile and conversational context, and then grounding the retrieval to it. (6) \textbf{MQ4CS}~\cite{lupart2024irlab}: Conducting multi-aspect query generation via GPT-4 and aggregating multiple result lists with a re-ranker. (7) \textbf{HumanQR}: Manual query rewriting annotations with personalization provided in the original datasets. 

For the second comparison category, we compare with four different personalized weight identification approaches in ranking fusion, including: (1) \textbf{Random Personalization}: the weight for personalization is random initialed between $0$ and $1$; (2) \textbf{Equal Personalization}: the weight for personalization is set the same as the non-personalized ones, which treat both of them equally; 
(3) \textbf{User-based Entropy}: 
Inferring the degree of personalization using user profile topic entropy~\cite{mei2008entropy,wang2013personalized}: different aspects in a user profile are treated as distinct topics. A higher topic entropy indicates that abundant user information is evenly distributed across these aspects, which in turn suggests a greater need for personalization.
(6) \textbf{DEPS}~\cite{liu2024personalize}: Estimating the need for personalization based on the diversity of candidate documents — that is, the greater the semantic differences among the documents, the higher the required level of personalization. 
See Sec.~\ref{sec: Implementation Details} for implementation details.

\subsection{Implementation Details}
\label{sec: Implementation Details}
\noindent \textbf{For \ours{}.} Our \ours{} framework integrates GPT-4o~\cite{gpt4o} for personalization level prediction and query reformulation, SPLADE (v3)~\cite{lassance2024splade} as the retriever, and monoT5~\cite{nogueira2020document} as the re-ranker.
BM25~\cite{robertson1995okapi}, ANCE~\cite{xiong2020approximate}, and RankGPT~\cite{sun2023chatgpt} are compared for an analysis of the choice of ranking models. 
All models use default hyperparameters.
Input query, pseudo response, and passage lengths are truncated to 64, 256, and 256 tokens, respectively.
The prompts to produce query reformulations and pseud-responses follows the RAR setting in  LLM4CS~\cite{mao2023large}.
Since our personalization-aware ranking fusion requires relevance judgments for fusion weight selection, we use one dataset for it and apply the weights to another for inference, e.g., weights selection on iKAT-23 and test on iKAT-24, vice versa.
Similarly, the example for both CoT and in-context learning employed in the prompt is randomly selected between two datasets.
Three query reformulation types -- the non-personalized query $q_n^{\prime}$ alone, with its potential answer $q_n^{\prime} \circ r_n^{\prime}$, and the personalized one $q_n^{u} \circ r_n^u$ would be obtained in each query turn in our implementation setting.
The estimated final fusion weights for corresponding reformulated queries with the three personalization levels are $(0.36, 0.17, 0.47)$, $(0.35, 0.2, 0.45)$, and $(0.25, 0.2, 0.55)$ for iKAT-23 and $(0.2, 0.38, 0.42)$, $(0.28, 0.36, 0.36)$, and $(0.23, 0.4, 0.37)$ for iKAT-24, respectively.  
More details can be found in our released code. 

\noindent \textbf{For Baselines.}
For the comparison with existing CQR systems, we implement each baseline method on our own, using the released checkpoints for T5QR and ConvGQR, or the original prompt for the remaining.
Besides, the number of top-$k$ passages used in the re-ranker is $100$ across all compared baselines and our \ours{}, except the MQ4CS, which re-ranks top-$1000*|Q|$ ($|Q|$ is the number of reformulated queries in each turn) following its original settings.
For the comparison with other approaches for personalized weight identification,
the \textit{Random Personalization} assigns a weight between $[0,1]$ to each $w_1$, $w_2$, and $w_3$, while the \textit{Equal Personalization} sets them with the same value.
About the \textit{User-based Entropy} method, for each sentence in the user profile $s_k \in \{\mathcal{U}\}_{k=1}^K$ encoded as $\bm{v}_k=\mathbf{E}(s_k)$, the resulting probability distribution over user aspects is defined as $p_k = \frac{1 - \frac{1}{K} \sum_{j=1}^K \bm{v}_k^\top \bm{v}_j}{\sum_{i=1}^K \left(1 - \frac{1}{K} \sum_{j=1}^K \bm{v}_i^\top \bm{v}_j\right)}$.
Then, the personalization need weight $w_3$ for the query is computed using the normalized Shannon entropy as $w_3 = -\frac{\sum_{k=1}^K p_k \log p_k}{\log K}$ based on this distribution.
For the \textit{DEPS} method, the personalized weight $w_3$ is decided by the aggregated representation of the retrieved candidate documents $\mathbf{E}(\mathcal{D}_{\text{non-p}}^{\text{agg}})$ from the non-personalized query as $w_3 = \sigma(||\mathbf{E}(q_n^u+r_n^u) - \mathbf{E}(\phi(\mathcal{D}_{\text{non-p}}^{\text{agg}}))||_2)$, where $\phi(\mathcal{D}_{\text{non-p}}^{\text{agg}})=\mathcal{R}(q_n^{\prime}+r_n^{\prime})$. 

\section{Experimental Results}

\subsection{Overall Performance}


To ensure fair and thorough comparison, we conduct open and aligned comparisons of our APCIR with baseline methods. The open comparison follows the instruction of TREC 
that any backbone model and external resource can be used, where the results are quoted from the original papers. The aligned comparison is conducted using fixed query rewriter (GPT-4o), retriever (BM25), and reranker (monoT5) among \ours{} and other main competitors.
The overall performance is shown in Table~\ref{table: main_results}, where can make the following observations:

(1) Our \ours~consistently outperforms all compared methods across two TREC iKAT datasets in both comparison settings and surpasses the human-rewritten queries (HumanQR) on the iKAT-24 dataset, demonstrating its superior effectiveness in personalized CIR.
The strong performance can be attributed to two aspects: i) explicit personalization level identification enables dynamic personalized results for each query turn compared to solely relying on LLM's implicit level of personalization, and ii) personalization-aware ranking fusion mechanism prevents over-personalization and provides abundant search results covering more aspects. 
We will further analyze these in Sec.~\ref{sec: ablation}.

(2) Comparing two evaluation settings, the use of different re-rankers has a significant impact on the final results. For example, GtR and our \ours~ have performance degradation while employing a re-ranker on iKAT-23 within the open comparison. However, the monoT5 re-ranker significantly improves the ranking results for each system in the aligned comparison.
Such a phenomenon indicates the importance of selecting a suitable re-ranker. For example, MQ4CS leverages SPLADE and DeBERTa \cite{lassance2023naver} for retrieval and re-ranking, where Deberta is fine-tuned with the hard negatives mined by SPLADE. We will provide more detailed analyses on the impact of using various combinations of search models in Sec.~\ref{sec: Impact of Search Models}.
In addition, we notice that MQ4CS, which also leverages the multiple query fusion paradigm by re-ranking the top-$1000*|Q|$ ($|Q|$ is the number of reformulated queries in each turn), may achieve better performance on some recall metrics than ours, which re-ranks the top-100. This might imply that increasing the size of the retrieval lists to be fused helps increase the aspects covered by different ranking lists, although at the cost of a higher latency.

(3) The HumanQR method underperforms our \ours~and MQ4CS on iKAT-24, while achieves the best result on iKAT-23. This indicates that human reformulation for personalized CIR does not necessarily produce the best search queries. Our method can be a good alternative to the costly human reformulation.  
\subsection{Comparison with other Personalized Weight Identification Methods}
\label{sec: weight comparison}
In this section, we further compare our \ours{} with other personalized weight identification methods to examine the effectiveness of our dynamically selected weights for ranking fusion. 
The comparison results are shown in Table~\ref{table: personalized_weight}.
We can observe that our method achieves the best performance and improves NDCG@3 scores by 2.9\% and 1.8\% absolute gain over the second-best method on two datasets, respectively.
These improvements demonstrate the superior capacity of our approach to identify the personalized needs for the given query in CIR.
Besides, the methods with specific personalized weight identified mechanisms consistently outperform the ones with random or no personalization, which emphasizes the importance of personalization needs.

\begin{table}[t]
    \centering
    \caption{Performance of \ours~ against other personalized weight identification methods for ranking fusion.}
    \vspace{-3ex}
    \scalebox{0.95}{
    \begin{tabular}{lcccc}
    \toprule
        \multicolumn{1}{c}{\multirow{2}[2]{*}{Method}} & \multicolumn{2}{c}{iKAT-23} & \multicolumn{2}{c}{iKAT-24} \\ 
        \cmidrule(lr){2-3} \cmidrule(lr){4-5}
        & MRR & N@3 & MRR & N@3 \\ 
        \midrule
        No Personalization & 46.0 & 20.6 & 66.8 & 37.1\\
        Random Personalization   & 42.2 & 20.3 & 73.2 & 39.6 \\
        Equal Personalization    & 53.6 & 28.2 & 81.2 & 49.0 \\
        User-Based Entropy       & 46.9 & 23.7 & 74.8 & 42.1 \\
        DEPS                     & 50.8 & 26.8 & 78.5 & 45.9 \\
       APCIR (Ours)       & \textbf{57.3}$^\dagger$ & \textbf{31.1}$^\dagger$ &  \textbf{82.7}$^\dagger$ & \textbf{50.8}$^\dagger$  \\
        \bottomrule
     \end{tabular}}
     \label{table: personalized_weight}
\vspace{-2ex}
\end{table}

\subsection{Ablation Study}
\label{sec: ablation}
\begin{figure}[t]
\centering
\includegraphics[width=0.9\linewidth]{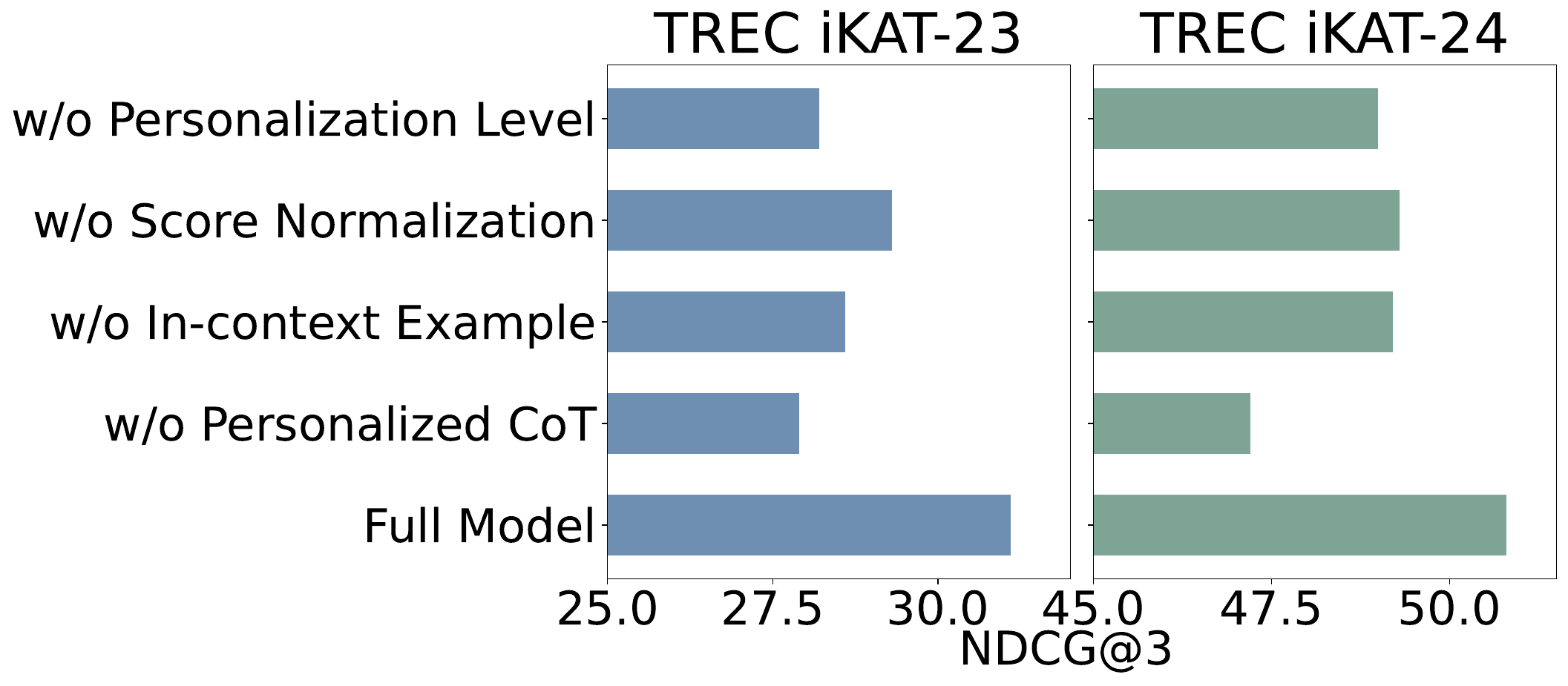}
\vspace{-3ex}
\caption{Ablation studies on the effectiveness of each component in our \ours~framework based on SPLADE model.}
\label{fig: ablation}
\vspace{-4ex}
\end{figure}
We conduct ablation studies to understand the effects of different components in our \ours~framework and analyze how ranking fusion can help avoid over-personalization. 

\noindent \textbf{Effects of Different Components.}
The impact of each component is shown in Figure~\ref{fig: ablation}.
We observe that removing each component would lead to performance drops, demonstrating the utility of each component. 
Among various mechanisms, the personalized CoT and personalization identification with explicit level contribute more than the others. 
The potential explanation is that personalized CoT may enhance the implicit reasoning of LLMs to judge the required level of personalization. Without human knowledge in CoT, 
LLMs may have difficulty in identifying the personalization need of a query. 
On the other hand, this confirms our conjecture that knowing the exact personalization level can help LLMs to extract more reasonable personal information from the user profile.

\noindent\textbf{Effect of Personalized Fusion against Over-Personalization.}
We compare the performance of the personalized and de-personalized queries in two settings to analyze how retrieval fusion overcomes query drift caused by introducing always personalized terms.
The de-personalization is done manually following the principle of removing the personalized part with minimum editing, thus reducing over-personalization.
For instance, the personalized query ``\textit{Can you help me find a suitable skincare routine, considering that I have oily, acne-prone, and dehydrated skin, and I prefer using Chanel products?}'' would be de-personalized as ``\textit{Can you help me find a suitable skincare routine?}''.
The results are shown in Table~\ref{table: ablation_perQ}.
We can see that personalized queries suffer from a performance drop due to over-personalization when used as a stand-alone query for search. However, using these seemingly worse personalized queries in our fine-grained fusion can outperform their de-personalized counterparts. 
This result demonstrates the ability of fusion to mitigate the effect of over-personalization. In fact, if personalization is considered to be unnecessary, the personalized retrieval may take a less important role in the final results through a lower fusion weight. Thus, fusion weight tuning for different personalization levels is an effective way to prevent over-personalization.
\begin{table}[t]
    \centering
    \caption{Ablation studies on the different usages of the reformulated personalized queries based on SPLADE model.}
    \vspace{-3ex}
    \scalebox{0.9}{
    \begin{tabular}{llcccc}
    \toprule
        \multicolumn{1}{c}{\multirow{2}[2]{*}{Query Usage}} & \multicolumn{1}{c}{\multirow{2}[2]{*}{Query Type}} & \multicolumn{2}{c}{iKAT-23} & \multicolumn{2}{c}{iKAT-24} \\ 
        \cmidrule(lr){3-4} \cmidrule(lr){5-6}
        & & MRR & N@3 & MRR & N@3 \\ 
        \midrule
        \multicolumn{1}{c}{\multirow{2}{*}{Stand-alone}} & Personalized Q & 46.8 & 23.4 & 71.5 & 40.6 \\
        & \quad + De-personalized         & \textbf{48.0} & \textbf{24.1} & \textbf{74.2} & \textbf{42.4} \\
        \midrule
       \multicolumn{1}{c}{\multirow{2}{*}{\makecell[c]{Fusion}}} & Personalized Q                         & \textbf{57.3} & \textbf{31.1} & \textbf{82.7} & \textbf{50.8} \\
        & \quad + De-personalized          & 54.4 & 29.3 & 82.5 & 49.4 \\
        \bottomrule
     \end{tabular}}
     \label{table: ablation_perQ}
\vspace{-2ex}
\end{table}

\subsection{Comparison of Retrieval Fusion Strategies}
Various other fusion strategies can achieve adaptive personalized CIR. In this section, we compare different fusion strategies, including reciprocal rank fusion, round-robin fusion, query representation fusion, and our proposed personalization-aware fusion approach.
The results are reported in Table~\ref{table: fusion}. We can see that our personalization-aware fusion approach outperforms the others, showing that it can achieve more adaptive personalized results. 
The reason is that the personalization-aware fusion can dynamically assign more appropriate weights to different queries according to the identified need for personalization, while the other strategies using fixed weights for all queries are unable to do it. 
In addition, our observations also suggest that fusing multiple reformulated queries with various personalization levels can achieve better results, due to discerning the necessary personalization requirement for the given query to prevent over-personalization.
Thus, more sophisticated mechanisms beyond the retrieval fusion paradigm can be explored later for more adaptive personalized CIR.

\begin{table}[t]
    \centering
    \caption{Impact of incorporating different retrieval fusion strategies with our \ours~based on SPLADE model.}
    \vspace{-3ex}
    \scalebox{0.95}{
    \begin{tabular}{lcccc}
    \toprule
        \multicolumn{1}{c}{\multirow{2}[2]{*}{Fusion Strategy}} & \multicolumn{2}{c}{iKAT-23} & \multicolumn{2}{c}{iKAT-24} \\ 
        \cmidrule(lr){2-3} \cmidrule(lr){4-5}
        & MRR & N@3 & MRR & N@3 \\ 
        \midrule
        Reciprocal Rank Fusion & 54.1 & 27.6 & 79.7 & 47.4 \\ 
        Round Robin Fusion          & 43.5 & 22.0 & 72.8 & 38.6 \\
        Query Representation Fusion & 40.0 & 19.0 & 70.3 & 37.9 \\
        Personalization-aware Fusion & \textbf{57.3} & \textbf{31.1} & \textbf{82.7} & \textbf{50.8} \\
        \bottomrule
     \end{tabular}}
     \label{table: fusion}
\vspace{-4ex}
\end{table}

\subsection{Impact of Using Different Search Models}
\label{sec: Impact of Search Models}
\begin{figure}[t]
\centering
\includegraphics[width=0.9\linewidth]{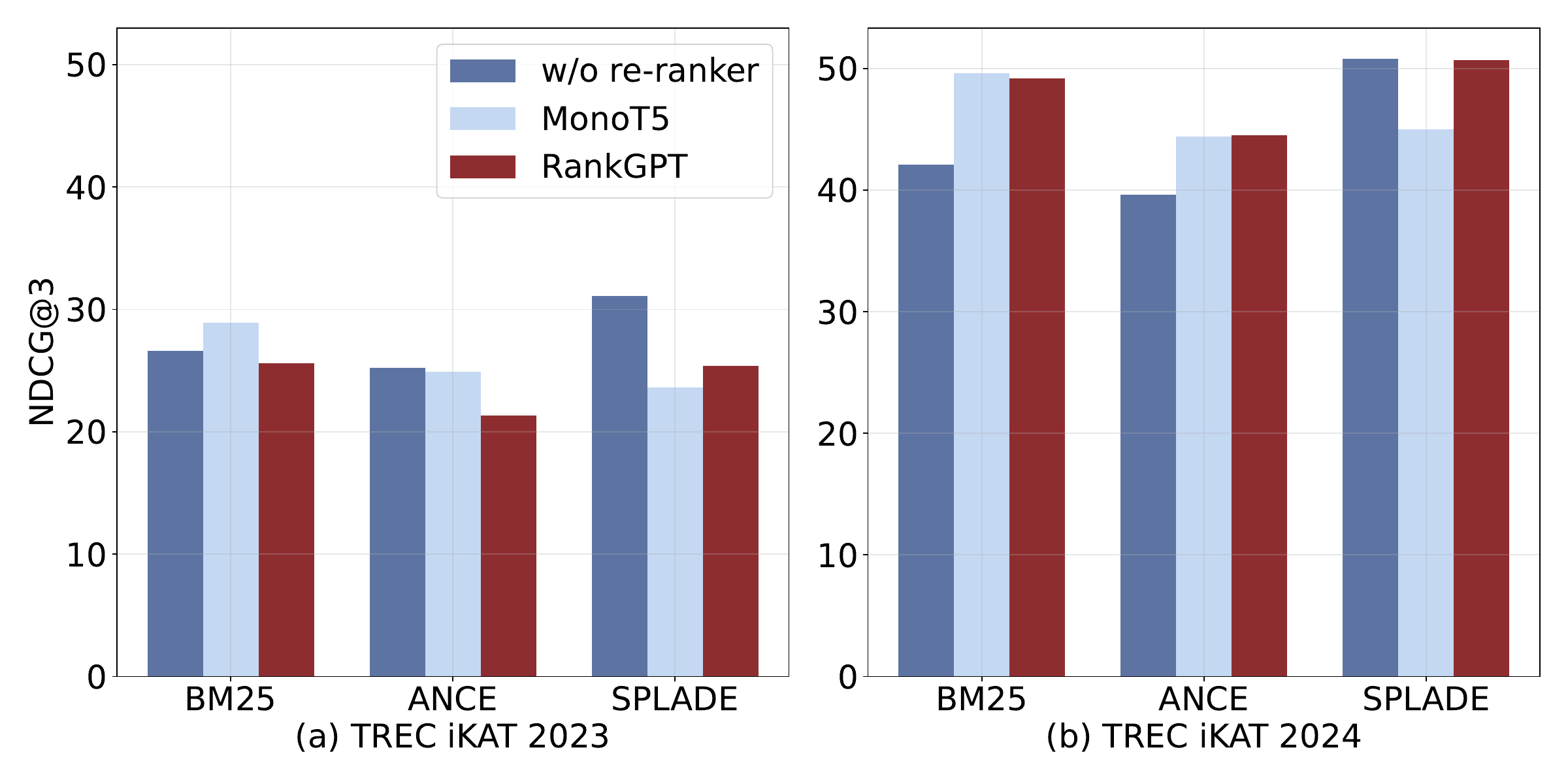}
\vspace{-4ex}
\caption{Impact of using different retrievers and re-rankers.}
\label{fig: reranker}
\vspace{-2.5ex}
\end{figure}

One of the advantages of our \ours~framework is that it can be adapted to any backbone model for retrieval and re-ranking. Thus, we conduct comprehensive experiments to verify the impact of various search models in \ours. 
Figure~\ref{fig: reranker} shows the effectiveness of \ours~with different types of retriever: sparse (BM25), dense (ANCE), hybrid (SPLADE), and two kinds of re-ranker: SLM-based (monoT5) and LLM-based (RankGPT).
We can see that the re-rankers are not always helpful. BM25 and ANCE benefit from the rerankers on iKAT 2024, while SPLADE does not. On iKAT 2023, the impact of re-rankers tends to be negative in general. 
Together with the results in Table~\ref{table: main_results}, these observations indicate that re-ranking might not be necessary for personalized CIR. 
For the first-stage retrieval,  SPLADE outperforms the other retrievers. The superior performance might be attributed to its abundant hybrid representation of text and emphasis on the importance of keywords through its sparse representation, which acts as a term-level de-noising for the long-form query in CIR~\cite{mao2023learning,lupart2024disco}.

\begin{table}[t]
    \centering
    \caption{Statistic of personalized requirement judgments based on LLM and Human, and their overlap in each group.}
    \vspace{-3ex}
    \scalebox{0.87}{
    \begin{tabular}{lccccccc}
    \toprule
    \multicolumn{1}{c}{\multirow{2}[2]{*}{Datasets}} & \multicolumn{3}{c}{APCIR Judgments} & \multicolumn{2}{c}{Human} & \multicolumn{2}{c}{Overlap} \\ 
    \cmidrule(lr){2-4} \cmidrule(lr){5-6} \cmidrule(lr){7-8}
    & \small Per. & \small Part. \small Per. & \small No Per. & \small Per. & \small No Per. & \small Per. & \small No Per.\\ 
    \midrule
    iKAT 23 & 23 & 85 & 68 & 63 & 113 & 77.8\% & 47.8\% \\
    iKAT 24 & 38 & 21 & 44 & 49 & 54  & 79.6\% & 63.0\% \\
    \bottomrule
     \end{tabular}}
     \label{table: per_level}
\vspace{-3ex}
\end{table}
\subsection{Analysis of Personalization Identification}
Table~\ref{table: per_level} shows the statistics of personalization requirement per query turn identified by our \ours~framework and the human annotations.
To estimate their overlap, we consider the personalized and partial personalized in LLM-based systems both as personalized in human-judged criteria.
The obvious gap in the judgments for both datasets indicates the potential annotation discrepancy in the original datasets, e.g., the annotated user information might not be useful for personalized retrieval, which is consistent with previous studies~\cite{mo2024leverage}. Indeed, given a query, its conversational context, and personal information, a human may not always judge correctly whether the personal information can help in retrieval.

Additionally, we conduct a retrieval evaluation grouped by the identified levels of personalization across various retrieval models. The performance, measured by NDCG@3 score, is presented in Table~\ref{table: per_level_NDCG}. The results reveal that queries requiring partial personalization are often the most challenging for retrieving relevant passages. 
Conversely, queries that do not require personalization are easier to deal with. This contrast further demonstrates the importance of determining the appropriate level of personalization for a query so that it can be personalized adequately.
\begin{table}[t]
    \centering
    \caption{Retrieval performance (NDCG@3) of queries with various personalization levels w.r.t different retrievers.}
    \vspace{-3ex}
    \scalebox{0.87}{
    \begin{tabular}{lcccccc}
    \toprule
    \multicolumn{1}{c}{\multirow{2}[2]{*}{Model}} & \multicolumn{3}{c}{iKAT-23} & \multicolumn{3}{c}{iKAT-24} \\ 
    \cmidrule(lr){2-4} \cmidrule(lr){5-7} 
    & Per. & Part. Per. & No Per. & Per. & Part. Per. & No Per.\\ 
    \midrule
    BM25 & \textbf{32.4} & 23.9 & 28.0 & 31.9 & 39.0 & 49.6\\
    ANCE & 21.1 & 22.7 & 30.0 & 36.0 & 36.0 & 41.4\\
    SPLADE & 29.5 & \textbf{29.6} & \textbf{33.6} & \textbf{47.2} & \textbf{43.3} & \textbf{58.9}\\
    \bottomrule
     \end{tabular}}
     \label{table: per_level_NDCG}
\vspace{-2ex}
\end{table}
\subsection{Impact of Query Reformulation Models}
We evaluate the generalization ability of our \ours~based on different LLMs for personalized query reformulation, including open-source and commercial models.
Table~\ref{table: QR models} presents the performance of various LLMs when explicit personalization identification and retrieval fusion are applied, compared to a method without them, i.e., asking LLMs to personalize the query through a single prompt. 
The results demonstrate that our approach consistently outperforms the baseline method across different backbone models, demonstrating the effectiveness and robustness of our personalized CIR approach using personalization detection and fusion, compared to a simple method using a single prompt asking LLMs to reformulate the query. 
We also see that this is true for any backbone LLMs. So, the effectiveness of our approach is independent from the LLMs used. 
\begin{table}[t]
    \centering
    \caption{Impact of using different backbone LLMs for personalized query reformulation on two settings.}
    \vspace{-3ex}
    \scalebox{0.95}{
    \begin{tabular}{lcccc}
    \toprule
        \multicolumn{1}{c}{\multirow{2}[2]{*}{LLM-QR Model}} & \multicolumn{2}{c}{iKAT-23} & \multicolumn{2}{c}{iKAT-24} \\ 
        \cmidrule(lr){2-3} \cmidrule(lr){4-5}
        & MRR & NDCG@3 & MRR & NDCG@3 \\ 
        \midrule
        \multicolumn{5}{c}{w/o explicit identification and fusion (Baseline)}\\
        \midrule
        GPT-4o &  43.1 & 22.6 & 71.9 & 40.0 \\
        ChatGPT-3.5 &  36.9 & 19.8 & 61.4 & 32.3 \\
        LLaMA-3.1-8B & 38.0 & 18.6 & 63.0 & 35.6 \\
        Mistral-2-7B & 35.9 & 18.4 & 58.9 & 33.8 \\
        \midrule
        \multicolumn{5}{c}{w/. explicit identification and fusion (Ours)}\\
        \midrule
        GPT-4o & \textbf{57.3} & \textbf{31.1} & \textbf{82.7} & \textbf{50.8} \\
        ChatGPT-3.5 & 48.8 & 25.1 & 74.2 & 44.9 \\
        LLaMA-3.1-8B & 50.5 & 26.8 & 79.9 & 44.5 \\
        Mistral-2-7B & 47.1 & 23.9 & 76.2 & 43.4 \\
        \bottomrule
     \end{tabular}}
     \label{table: QR models}
\vspace{-2ex}
\end{table}

\section{Conclusion and Future Work}
In this paper, we investigated the problem of adaptive personalization in conversational information retrieval.
We proposed a \ours~framework, which integrates explicit personalization identification and fusion of various types of query reformulation for each turn. We aim at a method that can incorporate different reformulated queries according to the identified need for personalization.
We showed in our experiments that such explicitly identified personalization level helps select the appropriate amount of personalization, thus 
prevent over-personalization. 
In addition, other types of query reformulation are also incorporated through fusion. The fusion weights trained for each level of personalization allow us to aggregate the ranking results of different forms of queries according to the required personalization level, which is shown to outperform other uniform fusion and personalized methods.

In future studies, it would be interesting to investigate a refined mechanism to determine better fusion weights for each given retrieval situation according to the available information and optimize the latency of the framework.
Besides, we could further explore the reasoning ability of LLMs to exploit the knowledge not present explicitly for personalized query reformulation.




\bibliographystyle{ACM-Reference-Format}
\bibliography{sample-base}

\appendix


\end{document}